\documentclass{article}

\newdimen\proofrulebreadth \proofrulebreadth=.05em
\newdimen\proofdotseparation \proofdotseparation=1.25ex
\newdimen\proofrulebaseline \proofrulebaseline=2ex
\newcount\proofdotnumber \proofdotnumber=3
\let\then\relax
\def\hfi{\hskip0pt plus.0001fil}
\mathchardef\squigto="3A3B
\newif\ifinsideprooftree\insideprooftreefalse
\newif\ifonleftofproofrule\onleftofproofrulefalse
\newif\ifproofdots\proofdotsfalse
\newif\ifdoubleproof\doubleprooffalse
\let\wereinproofbit\relax
\newdimen\shortenproofleft
\newdimen\shortenproofright
\newdimen\proofbelowshift
\newbox\proofabove
\newbox\proofbelow
\newbox\proofrulename
\def\shiftproofbelow{\let\next\relax\afterassignment\setshiftproofbelow\dimen0 }
\def\shiftproofbelowneg{\def\next{\multiply\dimen0 by-1 }%
\afterassignment\setshiftproofbelow\dimen0 }
\def\setshiftproofbelow{\next\proofbelowshift=\dimen0 }
\def\setproofrulebreadth{\proofrulebreadth}

\def\prooftree{
\ifnum  \lastpenalty=1
\then   \unpenalty
\else   \onleftofproofrulefalse
\fi
\ifonleftofproofrule
\else   \ifinsideprooftree
        \then   \hskip.5em plus1fil
        \fi
\fi
\bgroup
\setbox\proofbelow=\hbox{}\setbox\proofrulename=\hbox{}%
\let\justifies\proofover\let\leadsto\proofoverdots\let\Justifies\proofoverdbl
\let\using\proofusing\let\[\prooftree
\ifinsideprooftree\let\]\endprooftree\fi
\proofdotsfalse\doubleprooffalse
\let\thickness\setproofrulebreadth
\let\shiftright\shiftproofbelow \let\shift\shiftproofbelow
\let\shiftleft\shiftproofbelowneg
\let\ifwasinsideprooftree\ifinsideprooftree
\insideprooftreetrue
\setbox\proofabove=\hbox\bgroup$\displaystyle 
\let\wereinproofbit\prooftree
\shortenproofleft=0pt \shortenproofright=0pt \proofbelowshift=0pt
\onleftofproofruletrue\penalty1
}

\def\eproofbit{
\ifx    \wereinproofbit\prooftree
\then   \ifcase \lastpenalty
        \then   \shortenproofright=0pt  
        \or     \unpenalty\hfil         
        \or     \unpenalty\unskip       
        \else   \shortenproofright=0pt  
        \fi
\fi
\global\dimen0=\shortenproofleft
\global\dimen1=\shortenproofright
\global\dimen2=\proofrulebreadth
\global\dimen3=\proofbelowshift
\global\dimen4=\proofdotseparation
\global\count255=\proofdotnumber
$\egroup  
\shortenproofleft=\dimen0
\shortenproofright=\dimen1
\proofrulebreadth=\dimen2
\proofbelowshift=\dimen3
\proofdotseparation=\dimen4
\proofdotnumber=\count255
}

\def\proofover{
\eproofbit 
\setbox\proofbelow=\hbox\bgroup 
\let\wereinproofbit\proofover
$\displaystyle
}%
\def\proofoverdbl{
\eproofbit 
\doubleprooftrue
\setbox\proofbelow=\hbox\bgroup 
\let\wereinproofbit\proofoverdbl
$\displaystyle
}%
\def\proofoverdots{
\eproofbit 
\proofdotstrue
\setbox\proofbelow=\hbox\bgroup 
\let\wereinproofbit\proofoverdots
$\displaystyle
}%
\def\proofusing{
\eproofbit 
\setbox\proofrulename=\hbox\bgroup 
\let\wereinproofbit\proofusing
\kern0.3em$
}

\def\endprooftree{
\eproofbit 
  \dimen5 =0pt
\dimen0=\wd\proofabove \advance\dimen0-\shortenproofleft
\advance\dimen0-\shortenproofright
\dimen1=.5\dimen0 \advance\dimen1-.5\wd\proofbelow
\dimen4=\dimen1
\advance\dimen1\proofbelowshift \advance\dimen4-\proofbelowshift
\ifdim  \dimen1<0pt
\then   \advance\shortenproofleft\dimen1
        \advance\dimen0-\dimen1
        \dimen1=0pt
        \ifdim  \shortenproofleft<0pt
        \then   \setbox\proofabove=\hbox{%
                        \kern-\shortenproofleft\unhbox\proofabove}%
                \shortenproofleft=0pt
        \fi
\fi
\ifdim  \dimen4<0pt
\then   \advance\shortenproofright\dimen4
        \advance\dimen0-\dimen4
        \dimen4=0pt
\fi
\ifdim  \shortenproofright<\wd\proofrulename
\then   \shortenproofright=\wd\proofrulename
\fi
\dimen2=\shortenproofleft \advance\dimen2 by\dimen1
\dimen3=\shortenproofright\advance\dimen3 by\dimen4
\ifproofdots
\then
        \dimen6=\shortenproofleft \advance\dimen6 .5\dimen0
        \setbox1=\vbox to\proofdotseparation{\vss\hbox{$\cdot$}\vss}%
        \setbox0=\hbox{%
                \advance\dimen6-.5\wd1
                \kern\dimen6
                $\vcenter to\proofdotnumber\proofdotseparation
                        {\leaders\box1\vfill}$%
                \unhbox\proofrulename}%
\else   \dimen6=\fontdimen22\the\textfont2 
        \dimen7=\dimen6
        \advance\dimen6by.5\proofrulebreadth
        \advance\dimen7by-.5\proofrulebreadth
        \setbox0=\hbox{%
                \kern\shortenproofleft
                \ifdoubleproof
                \then   \hbox to\dimen0{%
                        $\mathsurround0pt\mathord=\mkern-6mu%
                        \cleaders\hbox{$\mkern-2mu=\mkern-2mu$}\hfill
                        \mkern-6mu\mathord=$}%
                \else   \vrule height\dimen6 depth-\dimen7 width\dimen0
                \fi
                \unhbox\proofrulename}%
        \ht0=\dimen6 \dp0=-\dimen7
\fi
\let\doll\relax
\ifwasinsideprooftree
\then   \let\VBOX\vbox
\else   \ifmmode\else$\let\doll=$\fi
        \let\VBOX\vcenter
\fi
\VBOX   {\baselineskip\proofrulebaseline \lineskip.2ex
        \expandafter\lineskiplimit\ifproofdots0ex\else-0.6ex\fi
        \hbox   spread\dimen5   {\hfi\unhbox\proofabove\hfi}%
        \hbox{\box0}%
        \hbox   {\kern\dimen2 \box\proofbelow}}\doll%
\global\dimen2=\dimen2
\global\dimen3=\dimen3
\egroup 
\ifonleftofproofrule
\then   \shortenproofleft=\dimen2
\fi
\shortenproofright=\dimen3
\onleftofproofrulefalse
\ifinsideprooftree
\then   \hskip.5em plus 1fil \penalty2
\fi
}

\usepackage{amsmath} 
\usepackage{amsthm}
\usepackage{amssymb}

\theoremstyle{plain}
\newtheorem{theorem}{Theorem}
\newtheorem{proposition}[theorem]{Proposition}

\newtheorem{lemma}[theorem]{Lemma}

\theoremstyle{definition}
\newtheorem{definition}[theorem]{Definition}

\newtheorem{examples}[theorem]{Examples}

\theoremstyle{remark}

\def\bF{\mathcal{F}}

\def\LPO{\mbox{\sc LPO}}
\def\MPO{\mbox{\sc MPO}}

\def\LPOL{\mbox{\sc LLPO}}

\def\SPR{\mbox{\sc SPR}}

\def\bS{\mbox{$\mathcal{S}$}}

\def\bF{\mbox{$\mathcal{D}$}}
\def\bData{\mbox{$\mathcal{C}$}}

\def\suc{\mbox{$\mathbf{c}$}}
\def\constante{\mbox{$\mathbf{b}$}}

\def\bVar{\mbox{$\mathcal{V}$}}

\def\valence{\nu}

\def\cbv#1{\mbox{$\ \vdash^{\!#1}\ $}}
\def\imp#1#2{\mbox{$\ \vdash^{\!#1}_{\!#2}\ $}}

\def\egalbF{\mbox{$\approx$}} 
\def\egalLite{\mbox{$=$}} 

\def\precbF{\mbox{$\prec_{\mathcal{D}}$}}
\def\precEqbF{\mbox{$\preceq_{\mathcal{D}}$}}
\def\NotprecEqbF{\mbox{$\not\preceq_{\mathcal{D}}$}}

\def\rpoEgal#1{\mbox{$\preceq_{#1}$}}
\def\rpoStrict#1{\mbox{$\prec_{#1}$}}

\def\rpoUn{\mbox{$\prec_{1}$}}
\def\NotrpoUn{\mbox{$\not\preceq_{1}$}}
\def\rpoUnEq{\mbox{$\preceq_{1}$}}

\def\lpoLite{\mbox{$\prec_{0}$}}
\def\lpoLiteEq{\mbox{$\preceq_{0}$}}

\def\lpo{\mbox{$\prec_{\mathrm{lpo}}$}}

\def\restrict#1#2{\mbox{$#1 \lceil #2$}}

\def\FV{\mathrm{Var}}

\def\val#1{\mbox{$[#1]$}}

\def\taille#1{\mbox{$|#1|$}}
\def\haut#1{\mbox{$\mathrm{ht}(#1)$}}

\def\redRefTrans{\mbox{${\stackrel{\tiny*\;}{\to}}$}}

\def\To{\to}
\def\transto{\mbox{${\stackrel{\mbox{\tiny $+\;$}}{\To}}$}}

\def\normto{\mbox{${\stackrel{\tiny !\;}{\To}}$}}
\def\reftransto{\mbox{${\stackrel{\tiny *\;}{\To}}$}}

\def\DHaut{\mbox{$\mathcal{DH}$}}

\def\sem#1{\mbox{$\{#1\}$}}
\def\T{\mbox{$\cal T$}}

\begin{document}

\title{The Light Lexicographic path Ordering}
\author{
E.A.~Cichon and J-Y.~Marion
\thanks{Loria, Projet Calligramme,
B.P.\ 239, 54506 Vand{\oe}uvre-l\`es-Nancy Cedex, France.
{\tt \{Adam.Cichon,Jean-Yves.Marion\}@loria.fr.}}
}

\maketitle

\begin{abstract}
We introduce syntactic restrictions of the lexicographic
path ordering to obtain the {\em Light Lexicographic Path
Ordering}. 
We show that the light lexicographic path
ordering leads to a characterisation of the functions computable in space
bounded by a polynomial in the size of the inputs. 
\end{abstract}

\section{Introduction}

Termination orderings have been particularly successful inventions for
proving the termination of rewrite systems. Their success is mainly due to
their ease of implementation and they are the principal tool used in modern
completion-based theorem provers. The two best
known termination orderings are the {\em multiset path ordering},
\MPO\footnote{Suppose that  $<_\Sigma$ is  an ordering on  the signature $\Sigma$.   The
{\em multiset  path ordering}, $<_{mpo}$,    on  $T(\Sigma)$
is      defined   recursively   as follows:
(i) $f(t_1,...,t_n) \leq_{mpo} t \To (\forall i \in 1..n ) t_i <_{mpo} t$,
(ii) $g <_\Sigma f$ and $(\forall i \in [1..m]) s_i <_{mpo}
f(t_1,\ldots,t_n)$ $\To$ $g(s_1,\ldots,s_m)  <_{mpo} f(t_1,\ldots,t_n)$,
(iii) $\{s_1,\ldots,s_m\} \ll_{mpo} \{t_1,\ldots,t_n\} \To
f(s_1,\ldots,s_m) <_{mpo} f(t_1,\ldots,t_n)$, where $\ll_{mpo}$   is  the multiset    ordering induced  by
$<_{mpo}$.}, and the {\em lexicographic path ordering},
\LPO\footnote{Suppose that  $<_\Sigma$ is  an ordering on  the signature $\Sigma$.   The
{\em lexicographic  path ordering}, $<_{lpo}$,    on  $T(\Sigma)$
is      defined   recursively   as follows:
(i) $f(t_1,...,t_n) \leq_{lpo} t \To (\forall i \in 1..n ) t_i <_{lpo} t$,
(ii) $g <_\Sigma f$ and $(\forall i \in [1..m]) s_i <_{lpo}
f(t_1,\ldots,t_n)$ $\To$ $g(s_1,\ldots,s_m)  <_{lpo} f(t_1,\ldots,t_n)$,
(iii) for some $i \in 1..n$, $s_1 = t_1, \ldots, s_{i-1} =
t_{i-1}, s_i <_{lpo} t_i$, and $s_{i+1} <_{lpo} t, \ldots, s_{m} <_{lpo} t$
$\To$ $f(s_1,\ldots,s_m) <_{lpo} f(t_1,\ldots,t_n)$.}, (and their variations or derivatives).

Termination orderings give rise to interesting theoretical questions
concerning the classes of rewrite algorithms for which they provide
termination proofs. It has been shown that $\MPO$ gives rise to a
characterisation of primitive recursion (see \cite{hof92}) and that $\LPO$
characterises the multiply recursive functions (see \cite{wei95}). While
both of these classes contain functions which are highly unfeasible, the
fact remains that many feasible algorithms can be successfully treated
using one or both of $\MPO$ or $\LPO$.

If we compare different syntactic characterisations of function classes we
observe that termination orderings can have a remarkable advantage over
other characterisations. Let us
compare $\MPO$ with {\em Strict Primitive
Recursion}, \SPR\footnote{The schemes for defining the primitive recursive functions over the 
natural numbers are as follows.
The {\bf Initial Functions} are 
\{ $Z(x) = 0$, 
$S(x) = x^\prime$, 
$U^n_i(x_0,\ldots,x_{n-1}) = x_i$ where $0\leq i\leq n-1$ \}.
The primitive recursive function are closed under (i) {\bf Composition}:
$f(\vec{x}) = g(h_1(\vec{x}),\ldots, h_m(\vec{x}))$, and (ii) 
{\bf Primitive Recursion}:
$f(\vec{x},0) = g(\vec{x})$, 
$f(\vec{x},y+1) = h(\vec{x},y,f(\vec{x},y))$.}. $\MPO$ is more general than
\SPR.
$\MPO$ is applicable to arbitrary equational specifications.
There is no {\em a priori} dependence on the data-type semantics. $\MPO$ always
proves termination of an \SPR~ programme over the tally numbers\footnote{ The
tally numbers are obtained using a single constant and a single, unary,
successor function symbol.}, as was observed by Plaisted \cite{Plai78}, but it
can also prove termination of other algorithms where the intended semantics are
not those of functions over the natural numbers. So $\MPO$ can be thought of as a
generalisation of \SPR~ which allows a broader class of
algorithms\footnote{$\MPO$ is intimately linked with the \SPR~ functions in the
following way: the computation time of an $\MPO$ algorithm is primitive recursive
in the size of the input.}.

Now, the results of Hofbauer and Weiermann, \cite{hof92, wei95},
indicate that $\LPO$ is considerably more powerful than $\MPO$ from the point
of view of computational complexity, and one might consider such extra
power to be redundant - one can show that $\LPO$ easily proves termination
of the Ackermann function\footnote{$A(0,y) = y+1$, $A(x+1, 0) = A(x, 1)$,
$A(x+1, y+1) = A(x, A(x+1,y))$} whereas no $\MPO$ termination proof for this
function is possible, but one criticism of this is to say ``so what, since
the Ackermann function is not feasibly computable?''. Yet, in some cases,
the builders of theorem provers have preferred \LPO. This is because of
\LPO's applicability to a wide range of {\em naturally arising algorithms}.
For example, we can paraphrase the Ackermann result stated above: for
syntactic reasons, $\MPO$ cannot prove termination of a simple feasible
function when its algorithm is based on a straightforward tail-recursion
whereas $\LPO$ will, in general, succeed. \\

In recent years, the lower complexity classes have been studied from a
syntactical point of view. Bellantoni and Cook \cite{BC92} restricted the
schemes for defining the primitive recursive functions to obtain a
syntactic characterisation of the functions computable in time bounded by a
polynomial in the size of the inputs. In parallel, Leivant \cite{Lei-1rep}
devised a method of data-tiering - a process which, roughly speaking,
assigns types to inputs and outputs in such a way as to restrict the class
of allowable algorithms within a given syntactic framework (strict
primitive recursion, for example).

In the present paper we introduce syntactic restrictions of the lexicographic
path ordering to obtain what we call the {\em Light Lexicographic Path
Ordering}. This is a follow-up to the work of Marion in
\cite{Marion00}\footnote{
We presuppose no familiarity with this work} where he
imposes syntactic restrictions on $\MPO$ to obtain a characterisation of the
polynomial time computable functions. We show that the light lexicographic path
ordering leads to a characterisation of the functions computable in space
bounded by a polynomial in the size of the inputs. The proof depends
essentially on the characterisation of the polynomial space computable
functions given in \cite{LM95}.
This caracterization relies on the ability
to capture recursion with parameter substitution,
as in~\cite{LM97,LM99}. \\


\section{First order functional programming}
The set of terms built up from a signature $\bS$ and  from a set of
variables $\bVar$ is $\T(\bS,\bVar)$.
A program, is defined by a quadruplet $\langle \to,\bData,\bF,f\rangle$ thus.
The set of data is the term algebra $\T(\bData)$ where
symbols in $\bData$ are called constructors. 
We shall always assume that $\bData$ contains, at least, a $0$-ary
constructor, i.e. a constant, and all other constructors are unary,
i.e. successors\footnote{The result presented can be extended
easily to lists. 
It might be possible to establish the same result 
on trees data structure by representing trees as directed acyclic
graphs. 
}.
For example, binary words will be represented by terms built up
from $\{\epsilon,s_0,s_1\}$ where $\epsilon$ is a constant denoting
the empty words, and $s_0$,$s_1$ are two successors.
$\bF$ is the set of function symbols of fixed arity $>0$. 
So, the full signature is
$\bS =\bData \cup \bF$. 
Rewrite rules are given by the binary relation $\to$.
Each rewrite rule is of the
form $g(p_1,\cdots,p_n) \to s$ where $g \in \bF$, the $p_i$'s
are patterns, that is terms  of $\T(\bData,\bVar)$,
and $s$ are terms of $\T(\bData \cup \bF,\bVar)$.
Moreover, $\FV(g(t_1,\cdots,t_n)) \subseteq \FV(s)$ where $\FV(t)$ is the
set of variables in a term $t$.
Lastly, the function symbol $f \in \bF$ is the main function symbol.

We define $u \to v$ to say that the term $v$ is obtained from $u$ by
applying a rewrite rule. The relation $\transto$ ($\reftransto$) denotes the
transitive (reflexive-transitive) closure of $\to$. 
We write $s \normto t$ to mean that 
$s \reftransto t$ and $t$ is in normal form.
A (ground) substitution $\sigma$ is a function from $\bVar$ into
$\T(\bData \cup \bF,\bVar)$ (resp. $\T(\bData \cup \bF)$).

Say that a program is confluent if the relation $\to$  is
confluent.
To give a semantic to programs, 
we just consider, as meaningful, the normal forms 
which are in the data set $\T(\bData)$. 

\begin{definition}
A confluent program $\langle\to,\bData,\bF,f\rangle$
computes the function 
$\sem{f}: \T(\bData)^{n} \mapsto \T(\bData)$
which is  defined as follows.
For all $u_1,\cdots,u_n \in \T(\bData)$, \\
  $\sem{f}(u_1,\cdots,u_n) = v$ if 
         $f(u_1,\cdots,u_n) \normto v$,
otherwise $\sem{f}(u_1,\cdots,u_n)$ is undefined.
\end{definition}

A program is terminating if there is no infinite derivation, that is
there is no infinite sequence of terms such that 
$t_0 \to t_1 \to t_2 \to \cdots$
One might consult~\cite{Der87} for a survey about rewriting
termination, and~\cite{DJ90} about general references on rewriting.

\section{Light Lexicographic Path Ordering}\label{sec:lpoLite}
We now describe our restriction of $\LPO$ which we call the {\em Light
Lexicographic Path Ordering} (\LPOL),

\begin{definition}
The {\em valency} of a function symbol $f$, of arity $n$, is a mapping
$\valence(f)\,:\,\{1,\cdots,n\} \mapsto \{0,1\}$.
We write  $\valence(f,i)$ to denote the valency of $f$ at its $i^{\rm th}$ argument.
\end{definition}

Valencies will allow us to combine two kinds of orderings to prove the
termination of a program. Valencies are to some extent related to the
Kamin-Levy~\cite{KM80} notion of functionals on orders. Indeed, a
functional on orderings can be defined as a status function on $\bF$ which
indicates how to compare terms, either in a lexicographic or in a multiset
way. Similarly, the valency of a function will also indicate how to compare
terms.

Above all, the notion of valency resembles the notion of data tiering which
was
introduced by Leivant in~\cite{Lei-1rep,Lei-predicativeI}.
The data tiering discipline ensures that the types of terms are also
tiered. Actually, function valencies are much more like normal and safe
position arguments as defined by Bellantoni and Cook
in~\cite{BC92}. Function valencies generalise this concept to functions
defined by means of recursive equations.
For the sake of readability, we use a notational convention similar to that
of~\cite{BC92}, and write $f(x_1,\cdots,x_n;y_1,\cdots,y_m)$, with a
semi-colon separating two lists of arguments, to indicate that 
$\valence(f,i)=1$ for $i \leq n$,
and $\valence(f,n+j)=0$ for $j \leq m$.
We shall write $f(\cdots,t_i,\cdots)$ to mean that the term
$t_i$ occurs at position $i$ in $f$, which is a position of valency
$\valence(f,i)$.

Throughout, we shall always assume that there is a precedence,
$\precEqbF$ on $\bF$ which is a total pre-order. 
As usual the strict precedence $\precbF$ is defined by $g \precbF f$
if $g \precEqbF f$ and $f \NotprecEqbF g$.
Also, the equivalence relation $\egalbF$ is  defined by $g \egalbF f$
if $g \precEqbF f$ and $f \precEqbF g$.
We also assume that $\egalbF$ respects the function symbol arities
and the valencies.
That is, if $f \egalbF g$ then the arity of $f$ and $g$ is $n$ 
and $\valence(f,i)=\valence(g,i)$ for all $i \leq n$.
On the other hand, the constructors of $\bData$ will be
incomparable, and that the only equivalence relation on $\bData$ will be
that of syntactic equality.

\begin{definition}
The partial ordering $\rpoUn$ on
 $\T(\bData \cup \bF,\bVar)$ is defined recursively  as follows:
\begin{enumerate}
\item If $s \rpoUnEq t$ and if $\suc \in \bData$, then $s \rpoUn
\suc(t)$.
\item If $\suc \in \bData$ and $s \rpoUn f(t_1,\cdots,t_m)$,
 then $\suc(s) \rpoUn f(t_1,\cdots,t_m)$.
\item If $s \rpoUnEq t_i$ and if $\valence(f,i)=1$
then $s \rpoUn f(\cdots,t_i,\cdots)$.
\item If $(g \precbF f)$ and if for all $i\leq n$,
$s_i \rpoUn f(t_1,\cdots,t_m)$,
then \\
$g(s_1,\cdots,s_n) \rpoUn f(t_1,\cdots,t_m)$.
\end{enumerate}
Here $\rpoUnEq = \rpoUn \cup \egalLite$, 
where $\egalLite$ is the syntactic identity,
and $f \in \bF$.
\end{definition}

\begin{examples}\
\begin{enumerate}
\item 
Two distinct terms of $\T(\bData)$ are incomparable by $\rpoUn$.
except if there are subterms.
\item $g(x,x;) \rpoUn f(x;)$ if $g \precbF f$.
\item $h(x;x) \rpoUn f(x;)$ if $h \precbF f$.
\item $f(x;) \NotrpoUn k(y;x)$ because $x$ is of valency $0$
in $k$.  
\item Two terms with the same root symbol are incomparable
with respect to $\rpoUn$. So $\rpoUn$ is not monotonic.
\end{enumerate}
\end{examples}

\begin{definition}
For each $f \in \bF$, we define
$$
\restrict{\bF}{f} = \{ g\ :\ g \precbF f\}
$$
\end{definition}

\begin{definition}
The partial ordering $\lpoLite$ on
$\T(\bData \cup\bF,\bVar)$ is defined recursively as follows:
\begin{enumerate}
\item If $s \lpoLiteEq t$ and if $f \in \bData \cup \bF$,
then $s \lpoLite f(\cdots,t,\cdots)$.
\item If $\suc \in \bData$ and $s \lpoLite f(t_1,\cdots,t_m)$,
 then $\suc(s) \lpoLite f(t_1,\cdots,t_m)$.
\item 
If $g \precbF f$ and for all $i\leq n$,
  $s_i \rpoStrict{\valence(g,i)} f(t_1,\cdots,t_m)$,\\
  then $g(s_1,\cdots,s_n) \lpoLite f(t_1,\cdots,t_m)$.
\item
If $g \egalbF f$ and if
$s_1 \egalLite t_1$, \ldots, $s_{p-1} \egalLite t_{p-1}$
and $s_{p} \rpoUn t_{p}$ where $\valence(f,p)=1$,
and for all $1 \leq j \leq n-p$, either
$s_{p+j} \rpoEgal{1} t_{p+j}$ and $\valence(f,p+j)=1$,
or $s_{p+j} \in \T(\bData \cup \restrict{\bF}{f},\bVar)$ and
$s_{p+j} \lpoLite f(t_1,\cdots,t_n)$ with
$\valence(f,p+j)=0$,
then $g(s_1,\cdots,s_n) \lpoLite f(t_1,\cdots,t_n)$.
\end{enumerate}
where $\lpoLiteEq = \lpoLite \cup \egalLite$, 
where $\egalLite$ is the syntactic identity,
and $f \in \bF$.
\end{definition}

The ordering $\lpoLite$ possesses the subterm property,
i.e. $t \lpoLite f(\cdots,t,\cdots)$
for all terms $t \in \T(\bData \cup \bF,\bVar)$ and $f \in \bData \cup \bF$.
The ordering $\lpoLite$ is monotonic with respect to arguments of
valency $1$, that is,  if
$s_i \rpoUn t_i$, where $\valence(f,i)=1$,  then
$f(\cdots,s_i,\cdots) \lpoLite f(\cdots,t_i,\cdots)$.

\begin{proposition}\label{prop:extension}
The ordering $\lpoLite$ is an extension of $\rpoUn$,
that is, if $s \rpoUn t$ then $s \lpoLite t$.
\end{proposition}


\begin{proposition}
The lexicographic path ordering $\lpo$ is an extension of $\lpoLite$,
that is if $s \lpoLite t$ then $s \lpo t$.
\end{proposition}

$\LPOL$ is the ordering formed by $(\rpoUnEq,\lpoLiteEq)$.
Termination follows from the fact that $\LPOL$ is a restriction
of $\LPO$.

\begin{definition}
An {\em $\LPOL$-program} is defined by a tuple
$\langle \to,\bData,\bF,f,\valence,\precbF \rangle$,
where $\langle\to,\bData,\bF,f\rangle$ is a confluent program,
$\valence$ is a valency function on $\bF$,
$\precEqbF$ is a precedence on $\bF$,
and such that 
for each rule $l \to r$, $r \lpoLite l$.
\end{definition}

\begin{theorem}\label{thm:lmpoTerminating}
Each $\LPOL$-program $\langle \to,\bF,\bData,f,\valence,\precbF \rangle$
 is terminating.
\end{theorem}

To use $\LPOL$ for a proof of termination, 
we must find an appropriate precedence $\precEqbF$
such that the program is terminating by $\LPO$.
Then, it remains to determine a correct 
valency function over $\bF$.

\begin{examples}\label{ex:lpoLite}\

\begin{enumerate}

\item Suppose that $\bData=\{\mathbf{\epsilon},\mathbf{0},\mathbf{1}\}$.

The set of constructor terms of $\T(\bData)$
represents binary words. The program below computes a function which
reverses a binary word.
\newcommand{\renv}{\mbox{\tt reverse}}
\begin{align*}
\renv(\mathbf{\epsilon};y) & \to y \\
\renv(\mathbf{i}(x);y) & \to \renv(x;\mathbf{i}(y))
\end{align*}
It is easy  to see that each rule is ordered by $\lpoLite$.
Notice that $\renv$ is defined by  a tail recursion whose termination
cannot be proved using \MPO.

\item \label{RecSP}
The following rules which define a recurrence with substitution of
  parameters.
This schema is important because it might
be seen as a template for simulating the computation of an alternating
Turing machine. Again, $\MPO$ does not prove the termination of this schema.
\begin{align*}
f(\constante;y)  & \to y \\
f(\suc(x);y) & \to h(x;y,f(x;\delta_0(;y)),f(x;\delta_1(;y)))
\end{align*}
By setting,
  $h \precbF f$ and $\delta_i \precbF f$,
the schema is ordered by $\LPOL$.
Indeed, $f(x;\delta_i(;y)) \lpoLite f(\suc(x);y)$
because (i) $\valence(f,2)=0$, (ii)
$\delta_i(;y) \in \T(\bData \cup \restrict{\bF}{f},\bVar)$
and (iii) $\delta_i(;y) \lpoLite f(\suc(x);y)$.
The ordering proof of the last equation is displayed below in sequent style.
$$
\prooftree
\[
 \[
 x=x
   \justifies
   x \rpoUn \suc(x)
 \]
\justifies x \rpoUn f(\suc(x);y)
\]
\[
y=y
\justifies y \lpoLite f(\suc(x);y)
\]
\[   
 \[ 
   x=x
   \justifies
   x \rpoUn \suc(x)
   \]
   \[
    \[
    y=y
    \justifies y \lpoLite f(x;y)
    \]
    \justifies \delta_i(;y) \lpoLite f(x;y)
   \]
\justifies f(x;\delta_i((;y)) \lpoLite f(\suc(x);y)
\]
\justifies h(x;y,f(x;\delta_0(;y)),f(x;\delta_1(;y))) \lpoLite f(\suc(x);y) 
\endprooftree 
$$
 
\item In this last exemple, we consider 
a program that computes $2^x+y$.
\begin{align*}
f(\constante,y)  & \to \suc(y) \\
f(\suc(x),y) & \to f(x,f(x,y))
\end{align*}
The termination of the above program cannot be shown by
$\lpoLite$. Indeed, if we use $\LPO$, 
we have to (i) compare lexicographically
$(x,f(x,y))$ with $(\suc(x),y)$
and 
then (ii) to show that $f(x,y)$ is less than 
$f(\suc(x),y)$. But 
$f(x,y) \not\in \T(\bData \cup \restrict{\bF}{f},\bVar)$
and so, this condition, which is imposed by $\lpoLite$,
does not hold. 
\end{enumerate}
\end{examples}

\section{Characterisation of Pspace}\label{sec:polyT}
Computational ressources, that is, time and space,
are measured relative to the size of the input arguments.
The size, $|t|$, of a term $t$ is the number of symbols in $t$.
$$
\taille{t} = \left\{
\begin{array}{lr}
1 & \mbox{if~} t \mbox{~is a constant} \\
\sum_{i=1}^n \taille{t_i} + 1 &\mbox{if~} t = f(t_1,\ldots,t_n)
\end{array}\right.
$$

A $\LPOL$-program $\langle \to,\bData,\bF,f,\valence,\precbF \rangle$
is computable in polynomial space,
if there is a Turing Machine $M$ such that
for each input $a_1,\cdots,a_n$ of $\T(\bData)$,
$M$ computes $\sem{f}(a_1,\cdots,a_n)$ in space
bounded by $P(\sum_{i\leq n} |a_i|)$.
We now state our main result.

\begin{theorem}\label{thm:det}
Each $\LPOL$-program is computable in polynomial space and,
conversely, each polynomial space function is computed by an
$\LPOL$-program.
\end{theorem}

The remain of the paper is devoted  to the proof of Theorem above.

\subsection{Poly-space functions are $\LPOL$-computable}

\begin{theorem}\label{thm:ML}
Each function $\phi$ which is computable in polynomial space is
represented by a $\LPOL$-program.
\end{theorem}

\begin{proof}
The proof is based on the characterisation of polynomial space
computable functions by means of ramified reccurence, reported
in~\cite{LM95}. We shall represent ramified functions by
$\LPOL$-programs.
Firstly, let us recall briefly how
ramified functions are specified.
Let $\bData$ be the set of constructors, and suppose that each
constructor is of arity $0$ or $1$.
Let $\bData_0,\bData_1,\cdots,\bData_k,\cdots$ be copies of the set
$\bData$. We shall say that $\bData_i$ is (a copy) of tier $i$.

A ramified function $F$ of arity $n$ must satisfy the following
condition :
The domain of a ramified function $F$ of arity $n$ is 
$\T(\bData_{i_1}) \times \cdots \times \T(\bData_{i_n})$, and
the range is $\T(\bData_{k})$ where the output tier is $k$
and $k = min_{j=1,n}(i_j)$.

The class of ramified functions is generated from constructors in
$\bData_i$, for all tiers $i$, and is closed under composition,
recursion with parameter substitution and flat recursion.

A ramified function $F$ is defined by flat recursion if
\begin{align*}
F(\constante,\vec{x}) = & G(\constante,\vec{x})
   && \text{where $\constante$ is a $0$-ary constructor in
   $\bData_{p}$} \\
F(\suc(t),\vec{x}) = & H(t,\vec{x})
 && \text{where $\suc$ is a unary constructor in $\bData_{p}$}
\end{align*}
where $\vec{x} = x_1,\cdots,x_n$
and $G$ and $H$ are defined previously. 
We see that the flat recursion template is ordered
by $\lpoLite$ because we can set $G,H \precbF F$.

A ramified function $F$ of output tier $k$
is defined by recursion with substitution of
parameters if for some $p > k$
\begin{align*}
F(\constante,\vec{x}) = & G(\vec{x})
 && \text{where $\constante$ is a $0$-ary constructor in $\bData_{p}$}
\\
F(\suc(t),\vec{x}) = & H(t,\vec{x},A_1,\cdots,A_m)
          && \text{where $\suc$ is a unary constructor in $\bData_{p}$}
\end{align*}
where  $A_j = F(t,\sigma_0^j(\vec{x}),\cdots,\sigma_n^j(\vec{x}))$.
The functions $G, H$ and $(\sigma_i^j)_{i \leq n;j \leq m}$ are ramified
functions.
The crucial requirement imposed on the ramified recursion schema
is that the output tier $k$ of $F$ be strictly
smaller than the tier $p$ of the recursion parameter. It follows that $p>0$.

Now the above template is ordered by $\lpoLite$ by
putting $F$ valencies thus.
\begin{align*}
\valence(F,p) & = 0 && \text{if the tier of the $p$\textit{th}
                                      argument  is $k$} \\
\valence(F,p) & = 1 && \text{otherwise}
\end{align*}
The termination proof is similar to the proof of
example~\ref{ex:lpoLite}(\ref{RecSP}).

We conclude that each ramified function is computable by an
$\LPOL$-program,
and so, following~\cite{LM95}, each polynomial-space computable function
is represented by an $\LPOL$-program.
\end{proof}

\subsection{$\LPOL$-programs are Poly-space computable}
The height $\haut{t}$ of a term $t$ is defined as the length
 of the longest branch in the tree $t$.
 $$
 \haut{t} = \left\{
 \begin{array}{lr}
 1 & \mbox{if } t \mbox{ is a constant or a variable} \\
 \max_{i=1}^n \haut{t_i} + 1 &\mbox{if } t = f(t_1,\ldots,t_n)
 \end{array}\right.
 $$

\begin{theorem}\label{thm:bounded}
Let  $\langle \to,\bData,\bF,f,\valence,\precbF \rangle$
be a $\LPOL$-program.
There is a polynomial $P$ such that
for all inputs $a_1,\cdots,a_n \in \T(\bData)$,
we have
\begin{enumerate}
\item \label{item:hauteur}
If $f(x_1,\cdots,x_n)\sigma \redRefTrans t$
then $\haut{t} \leq P(\sum_{i \leq n} |a_i)$.
\item \label{item:normal}
If $f(x_1,\cdots,x_n)\sigma \redRefTrans u$ and
$u \in \T(\bData)$ is a subterm of $t$, \\
then $|u| \leq P(\sum_{i\leq n} |a_i|)$.
\end{enumerate}
\end{theorem}

\begin{proof}
(1) is a consequence of Theorem~\ref{thm:poly},
of Lemma~\ref{lem:polyf} and of Lemma~\ref{lem:hauteurBornee}.
(2) is a consequence of (1) by observing that 
for each $u \in \T(\bData)$, we have  $\taille{u} = \haut{u}$.
\end{proof}

\begin{theorem}\label{thm:LPOL-PSPACE}
Let $\langle \to,\bData,\bF,f,\valence,\precbF \rangle$ be
an $\LPOL$-program.
For all inputs $a_1,\cdots,a_n \in \T(\bData)$,
the computation of
$f(a_1,\cdots,a_n) \normto v$, where $v\in\T(\bData)$,
is performed in
space bounded by $P(\sum_{i\leq n}|a_i|)$, where $P$ is some polynomial
which depends on the program.
\end{theorem}

\begin{proof}
Given a program $\langle \to,\bData,\bF,f,\valence,\precbF \rangle$,
the operational semantics of call by value are provided by a relation
$\cbv{\sigma} \subseteq \T(\bData \cup \bF,\bVar) \times \T(\bData)$
where $\sigma$ is a substitution over $\T(\bData)$.
The relation $\cbv{\sigma}$ is defined as the union of the family
$\imp{\sigma}{h}$ defined below :
\begin{itemize}
\item $x \imp{\sigma}{h} \sigma(x)$, if $x \in \bVar$ and
   $h = \taille{\sigma(x)}$.
\item $\constante \imp{\sigma}{1} \constante$, if $\constante$ is a
  $0$-ary constant of $\bData \cup \bF$.
\item $\suc(t) \imp{\sigma}{h+1} \suc(u)$ if $t \imp{\sigma}{h} u$
\item $f(t_1,\cdots,t_n) \imp{\sigma}{h} u$,
   if $t_1 \imp{\sigma}{h_1} u_1$,\ldots,$t_n \imp{\sigma}{h_n} u_n$,
   and $s \imp{\theta}{h_0} u$ \\
   where $f(u_1,\cdots,u_n)=t\theta$
   $t \to s$ is a rulw, and $h = \max_{i=1,n}(h_i + 1, h_0)$.
\end{itemize}
and $\cbv{\sigma} = \cup_{h \geq 0} \imp{\sigma}{h}$.
It is routine to verify that
$t\sigma  \normto v$
iff
$t \cbv{\sigma} v$, where $v \in \T(\bData)$.

The rules of the operational semantics described above form a
recursive
algorithm which is an interpreter of $\LPOL$-programs.
Put $\sigma_0(x_i)=a_i$.
The computation of $f(a_1,\cdots,a_n)$
consists in determining $u$ such that
 $f(x_1,\cdots,x_n) \imp{\sigma_0}{h} u$
for some $h$. 
Actually, $h$ is the height of the computation tree.

It remains to show that this evaluation procedure runs in
 space bounded by a polynomial in the
sum of sizes of the inputs.

For this, put $\DHaut(t) = \max\{ \haut{u}\ :\ t \reftransto u\}$.
By induction on $h$, we can establish that
 $t \imp{\sigma}{h} u$ implies $h \leq \DHaut(t\sigma)$.
As an immediate consequence of 
Theorem~\ref{thm:bounded}(\ref{item:hauteur}), 
we have $h \leq P(\sum_{i \leq n} |\sigma_0(x_i)|)$.

Now, at any stage of the evaluation,
the number of variables assigned by a substitution
is less or equal to (the maximal arity of a function symbol
$\times h$). 
Next, the size of the value of each variable is bounded
$P(\sum_{i \leq n} |\sigma_0(x_i)|)$ because 
 of Theorem~\ref{thm:bounded}(\ref{item:normal}).
Consequently, the space required to store a substitution is always less
than
$O(P(\sum_{i \leq n} |\sigma_0(x_i)|)^2)$.
We conclude that the whole runspace is bounded by
$O(P(\sum_{i \leq n} |\sigma_0(x_i)|)^3)$.
\end{proof}

\section{The bounding theorem}\label{sec:bound}
The purpose of this section is to prove Theorem~\ref{thm:bounded}.
We introduce a polynomial quasi-interpretation of 
 $\LPOL$-programs parameterised by $d \geq 2$. 
Given a $\LPOL$-program
$\langle \to,\bData,\bF,f,\valence,\precbF \rangle$.
we take $d$ such that for each rule $l \to r$, we have $d > |r|$.

The quasi-interpretation is given by a sequence of polynomials.
\begin{align*}
 F_0(X) & = X^d \\
 F_{k+1}(X) & = F_{k}^{d}(X)
\end{align*}
where $F_{k}^{a}(X)$ means $a$ iterations of $F_{k}$, i.e.
 $F_k(\cdots(F_k(X))\cdots)$.

Define $\bF_1,\cdots,\bF_k$ as the partition of $\bF$ determined by 
$\precEqbF$ such that  $g \in \bF_q$ and $f \in \bF_{q+1}$ iff
$g \precbF f$, and $f \egalbF g$ iff $f$ and $g$ are in $\bF_q$. 
We say that if $f$ is in $\bF_q$ then $f$ is of {\em rank} $q$.
Intuitively, we consider constructors as symbols of rank $0$. 
Now, the interpretation $\val{\ }$ on terms of $\T(\bData \cup \bF)$
is defined as follows.
\begin{itemize}
\item $\val{\constante}=d$ for every  constant $\constante$ of
$\bData$.
\item $\val{\suc(t)}= \val{t}+d$
 for every unary constructor $\suc$ of $\bData$.
\item $\val{f(t_1,\cdots,t_n)} =
 F_{k}(\max(d,\sum_{\valence(f,i)=1} \val{t_i})) 
                 + \max_{\valence(f,i)=0}(\val{t_i})$
for every $f \in \bF$ of rank $k$.
\end{itemize}

\begin{lemma}\label{lem:ProprieteInterpretation}\
\begin{enumerate}
\item \label{InterpreationMonotone}
 If $\val{s} \leq \val{t}$ then
 $\val{f(\cdots,s,\cdots)} \leq \val{f(\cdots,t,\cdots)}$.
\item \label{InterpretationSousterme}
 $\val{t} \leq \val{f(\cdots t \cdots)}$.
\end{enumerate}
\end{lemma}

\begin{theorem}\label{thm:poly}
Let  $\langle \to,\bData,\bF,f,\valence,\precbF \rangle$
be a an $\LPOL$-program.
For every substitution $\sigma$,
if $v\sigma \reftransto u\sigma$ then $\val{u\sigma} \leq
\val{v\sigma}$.
\end{theorem}

\begin{proof}
The demonstration of the theorem above is tedious.
The theorem is a consequence 
of Theorem~\ref{thm:Interpretation}
whose proof is detailed in the three next subsections.
For each ground substitution $\sigma$ and rule $l \to r$,
we have $\val{l\sigma} > \val{r\sigma}$ by
Theorem~\ref{thm:Interpretation} below.
The result follows from the monotonicity of the interpretation as
stated in
Lemma~\ref{lem:ProprieteInterpretation}(\ref{InterpreationMonotone}).
\end{proof}

\begin{lemma}\label{lem:polyf}
Let  $\langle \to,\bData,\bF,f,\valence,\precbF \rangle$
be a $\LPOL$-program. \\
for all inputs $a_1,\cdots,a_n \in \T(\bData)$,
  \begin{eqnarray} \label{eq:Polynome}
   \val{f(a_1,\cdots,a_n)} & \leq & P(\sum_{i\leq n} |a_i|)
  \end{eqnarray}
\end{lemma}

\begin{proof}
      Suppose that $f$ is a function symbol of rank $k$.
      Put $P(X) = F_k(d \cdot X) + d \cdot X$.
      By definition, we have $\val{f(x_1,\cdots,x_n)\sigma} =
      F_k(\sum_{\valence(f,i)=1} \val{a_i}) +
      max_{\valence(f,i)=0}(\val{a_i})$.
      Thus,
      $\val{f(x_1,\cdots,x_n)\sigma} \leq P(\sum_{i \leq n}
|a_i|)$, since $\val{a_i}=|a_i|$.
\end{proof}

\begin{lemma} \label{lem:hauteurBornee}
   $\haut{t} \leq \val{t}$
\end{lemma}

\begin{proof}
Straightforward by induction on $\haut{t}$.
\end{proof}

\subsection{Properties of $F_k$}
\begin{proposition}\label{prop:genFk}
For all $X$, $k$ and $d\geq 2$,
\begin{enumerate}
\item \label{exact} $F_k(X) = X^{d^{d^k}}$ 
\item \label{exposant}  $F_{k+1}^{\alpha}(X) = F_k^{\alpha \cdot d}(X)$
\item \label{croissanceSuite}  For all $j \leq k$ and $X \leq Y$,
we have $F_j(X) \leq F_k(Y)$ 
\item For all $X \geq d$ and $\alpha,\beta > 0$,
$F_k^{\alpha}(X) + F_k^{\beta}(X) \leq
                F_{k}^{\alpha+\beta}(X)$ \label{somme-composition}
\item  For all $X$ and $\alpha < d$, \label{thm-argument}
        $F_k^{\alpha}(X+1)+F_{k+1}(X) \leq F_{k+1}(X+1)$
\label{recfk}
\end{enumerate}
\end{proposition}

\subsection{Bounds on arguments of valency $1$}

\begin{lemma}\label{lem:rpoUn}
Let $s$ and $t=f(t_1,\cdots,t_n)$
be two terms of $\T(\bData \cup \bF,\bVar)$ such that $\FV(s)
\subseteq \FV(t)$.
Suppose that $s \rpoUn t$.
Let $\sigma$ be a ground substitution.

Assume  that  for all terms $u$,
$u \rpoUnEq t_i$ implies $\val{u\sigma} \leq \val{t_i\sigma}$,
for each $i \leq n$.
Then,
$\val{s\sigma}\leq F_k^{|s|}(A)$
where $f$ is of rank $k+1$
and $A = \max(d,\sum_{\valence(f,i)=1} \val{t_i\sigma})$.
\end{lemma}

\begin{proof}
The proof is by induction on $|s|$.
\begin{trivlist}
\item Assume $|s|=1$.

\item 
Suppose that $s$ is a constant of $\bData$.
We have $\val{s\sigma} = d \leq F_k^{|s|}(A)$,
by Proposition~\ref{prop:genFk}(\ref{croissanceSuite}).

\item 
Suppose that $s$ is a variable of $\bVar$.
We have $s \rpoUnEq t_i$ for some $i$ satisfying
$\valence(f,i)=1$ and $s \in \FV(t_i)$.
By the hypotheses of the lemma, $\val{s\sigma} \leq \val{t_i\sigma}$.
So, $s\sigma \leq F_k(A)$
by Proposition~\ref{prop:genFk}(\ref{croissanceSuite}).

\item Assume $|s|>1$.
\item 
Suppose that $s \rpoUnEq t_i$ and that $\valence(f,i)=1$.
Again, by the hypotheses of the lemma,
$\val{s\sigma} \leq \val{t_i\sigma}$ where $\valence(f,i)=1$.
So, $\val{s\sigma} \leq F_k^{|s|}(A)$
by Proposition~\ref{prop:genFk}(\ref{croissanceSuite}).

\item 
Suppose $s = g(s_1,\cdots,s_m)$ where $g$ is a function symbol of
$\bF$ of rank $j \leq k$ and
 that for all $i \leq m$,
 $s_i \rpoUn t$.
We have
\begin{align*}
\val{s\sigma} & = F_j(\sum_{i \leq m} \val{s_i\sigma}) && \text{by
            definition}\\
 & \leq F_j(\sum_{i \leq m} F_k^{|s_i|}(A)) 
 && \text{by induction hypothesis and by 
 Prop.~\ref{prop:genFk}(\ref{croissanceSuite})} \\
 & \leq F_j(F_k^{\sum_{i \leq m}|s_i|}(A)) 
 && \text{by Prop.~\ref{prop:genFk}(\ref{somme-composition})}
\\
 & \leq F_k^{\sum_{i \leq m}|s_i|+1}(A) && \text{by 
 Prop.~\ref{prop:genFk}(\ref{croissanceSuite})} 
\end{align*}
Lastly, the case when $s=\suc(s')$ where $\suc$ is a constructor
of $\bData$, is similar to the previous one, and so we skip it.
\end{trivlist}
\end{proof}

\subsection{An upper bound on the interpretation}

\begin{lemma}\label{lem:lpoLiteBorne}
Let $s$ and $t=f(t_1,\cdots,t_n)$
be two terms of $\T(\bData \cup \bF,\bVar)$
such that $\FV(s) \subseteq \FV(t)$
and $s \in \T(\bData \cup \restrict{\bF}{f},\bVar)$.
Suppose that $s \lpoLite t$.
Let $\sigma$ be a ground substitution.\\
Assume also, that for all $i \leq n$, and all terms $u$,
$u \lpoLiteEq t_i$ implies $\val{u\sigma} \leq \val{t_i\sigma}$.
Then,
\begin{equation}\label{eq1}
\val{s\sigma} \leq F_k^{|s|}(A)
 + \max_{\valence(f,i)=0}(\val{t_i\sigma})
\end{equation}
where $A = \max(d,\sum_{\valence(f,i)=1}\val{t_i\sigma})$ and $f$ is a
function symbol of rank $k+1$.
\end{lemma}

\begin{proof}
By induction on $|s|$.
\begin{trivlist}
\item Assume $|s|=1$.
\item 
Suppose that $s$ is the constant $\constante \in \bData$.
 We obtain
$\val{s\sigma} = \val{\constante} = d \leq F_k(A)$,
by~\ref{prop:genFk}(\ref{croissanceSuite}) and
so (\ref{eq1}) holds.
\item
Suppose that $s$ is a variable.
 We have $s \lpoLite t_i$ for some $i$.
By the hypotheses of the lemma we have
 $\val{s\sigma}\leq \val{t_i\sigma}$.
Therefore $\val{s\sigma} \leq \max_{\valence(f,i)=0}(\val{t_i\sigma})$.

\item Assume $|s|>1$.
Suppose that $s \lpoLiteEq t_i$.
The hypotheses of the lemma give $\val{s\sigma} \leq \val{t_i\sigma}$.
So (\ref{eq1}) holds.

\item
Suppose that $s = f_j(s_1,\cdots,s_m)$ where $f_j$ is a function symbol
of
$\bF$ of rank $j \leq k$.
Then, for all $i \leq m$, if $\valence(f,i)=1$,
we have $s_i \rpoUnEq t$.
Lemma~\ref{lem:rpoUn} yields $\val{s_i\sigma} \leq F_k^{|s_i|}(A)$,
so we have
\begin{align*}
F_j(\sum_{\valence(f_j,i)=1} \val{s_i\sigma})
 & \leq F_j(\sum_{\valence(f_j,i)=1} F_k^{|s_i|}(A))
 && \text{repl. $\val{s_i\sigma}$ by $F_k^{|s_i|}(A)$} \\
 & \leq F_j(F_k^{\sum_{\valence(f_j,i)=1} |s_i|}(A))
 && \text{by (\ref{prop:genFk}.\ref{somme-composition})} \\
 & \leq F_k^{1+\sum_{\valence(f_j,i)=1} |s_i|}(A)
 && \text{by 
(\ref{prop:genFk}.\ref{croissanceSuite})}
\end{align*}
Otherwise, $\valence(f,i)=0$ and we have $s_i \lpoLiteEq t$.
It follows by the induction hypothesis and
by monotonicity of $F_k$ that
$$
\max_{\valence(f_j,i)=0}(\max(d,\val{s_i\sigma}))
 \leq F_k^{\sum_{\valence(f,i)=0} |s_i|}(A)
 + \max_{\valence(f,i)=0}(\val{t_i\sigma})
$$
By definition, $\val{s\sigma} = F_j(\sum_{\valence(f_j,i)=1
}\val{s_i\sigma})
 + \max_{\valence(f_j,i)=0}(\val{s_i\sigma})$.
From the above inequalities, we get the following
bound on
$$
\val{s\sigma}\leq F_k^{1+\sum_{i \leq m} |s_i|}(A)
 + \max_{\valence(f,i)=0}(\val{t_i\sigma})
$$
The case when $s = \suc(s')$ is similar to the case above.

\end{trivlist}
\end{proof}

\begin{lemma}\label{lem:lpoLite}
Let $s$ and $t=f(t_1,\cdots,t_n)$
be two terms of $\T(\bData \cup \bF,\bVar)$
such that $\FV(s) \subseteq \FV(t)$.
Suppose that $s \lpoLite t$.
Let $\sigma$ be a substitution which assigns to each variable of $\FV(t)$
a ground term of $\T(\bData \cup \bF)$.
Assume also, that for all $i \leq n$, and all terms $u$,
$u \lpoLiteEq t_i$ implies $\val{u\sigma} \leq \val{t_i\sigma}$.
Then,
 $$\val{s\sigma} \leq F_k^{|s|}(A)
 + F_{k+1}(A-1)
 + \max_{\valence(f,i)=0}(\val{t_i\sigma})$$
where $A = \max(d,\sum_{\valence(f,i)=1}\val{t_i\sigma})$ and $f$ is a
function symbol of rank $k+1$.
\end{lemma}

\begin{proof}
The proof goes by induction on $|s|$.
However all the cases are the same as in  
 in lemma \ref{lem:lpoLiteBorne}, except the following one.

Suppose that $s = g(s_1,\cdots,s_m)$
where $g$ is a function symbol of $\bF$ of rank $k+1$.

We have $s_i \rpoUnEq t_i$ for each
position $i$ such that  $\valence(f,i)=1$.
Now, $s_i \rpoUnEq t_i$ implies $s_i \lpoLiteEq t_i$,
by Proposition~\ref{prop:extension}. 
So we can apply the lemma hypothesis,
and we obtain
$\sum_{\valence(g,i)=1} \val{s_i\sigma}
 < \sum_{\valence(f,i)=1} \val{t_i\sigma} = A$.
The inequality is strict because 
$s_p \rpoUn t_p$ for at least one $p$ such that 
$\valence(f,p)=1$.

On the other hand, if $\valence(f,j)=0$
we know that 
 $s_j \in \T(\bData \cup \restrict{\bF}{f},\bVar)$.
By Lemma~\ref{lem:lpoLiteBorne},
$\val{s_j\sigma} \leq
F_k^{|s_j|}(\sum_{\valence(f,i)=1}(\val{t_i\sigma}))
 + \max_{\valence(f,i)=0}(\val{t_i\sigma}) $.
Consequently,
$$
\max_{\valence(f,j)=0}(\val{s_j\sigma})
 \leq F_k^{\sum_{\valence(f,j)=0}|s_j|}(A) +
 \max_{\valence(f,j)=0}(\val{t_j\sigma})
$$

From both former inequalities, we conclude that
\begin{align*}
\val{s\sigma} & \leq F_{k+1}(A-1)
 +
F_{k}^{\sum_{\valence(f,j)=0}|s_j|}(A)
 + \max_{\valence(f,i)=0}(\val{t_i\sigma}) 
\end{align*}
\end{proof}

\begin{theorem}\label{thm:Interpretation}
Let $s$ and $t$ be two terms of $\T(\bData \cup \bF,\bVar)$
such that $|s|<d$. For every ground substitution $\sigma$,
 if $s \lpoLite t$ then $\val{s\sigma} < \val{t\sigma}$.
\end{theorem}

\begin{proof}
\label{pf:Interpretation}
The proof is by induction on $|t|$.
We have $|t|>1$.
Assume that $t=\suc(t')$, $\suc \in \bData$.
We have $s \lpoLiteEq t'$,
and so $\val{s\sigma} \leq \val{t'\sigma}$ by
induction hypothesis.
Assume that $t=f(t_1,\cdots,t_n)$.
Lemma~\ref{lem:lpoLite} yields that
$\val{s\sigma} \leq F_k^{|s|}(A)
 + F_{k+1}(A-1)
 + \max_{\valence(f,i)=0}(\val{t_i\sigma})$.
By Proposition~\ref{prop:genFk},
 $F_k^{|s|}(A) + F_{k+1}(A-1) < F_{k+1}(A)$.
Therefore, 
$\val{s\sigma} < F_{k+1}(A) + \max_{\valence(f,i)=0}(\val{t_i\sigma})
 = \val{f(t_1,\cdots,t_n)\sigma}$.
\end{proof}

\section{conclusion}
The result gives a purely syntactic characterisation
of functions computed in polynomial space by
mean of programs interpreted over infinite domains.
There are  few such characterisations like Leivant-Marion~\cite{LM95,LM97},
Leivant~\cite{LeivantCSL99}, and Jones~\cite{Jones00}. 
The other characterisations of functions computed in polynomial space 
deal with finite model theory or with bounded recursions.

Actually, we think that the main interest of this result lies on the
fact that we have illustrated that
the notion of valency and of termination ordering
is a tool to analyse programs. 
Putting things quickly, the role of valencies is to predict 
argument behaviour, and the role of termination ordering is
to capture algorithmic patterns. 

Finally, it is worth noticing that a consequence of this study
and of the recent work~\cite{MM00} is the following : 
A program which terminates by $\LPO$ and admits a quasi
interpretation bounded by a polynomial runs in polynomial space,
and conversely.

\end{document}